\documentclass[12pt]{iopart}

\usepackage{graphicx}
\usepackage{bm}

\begin{document}
\title[Electronic structure of GdN, and the influence
of exact exchange]{Electronic 
structure of GdN, and the influence
of exact exchange}

\author{K. Doll}

\address{Max-Planck-Institute for Solid State Research,
Heisenbergstra{\ss}e 1, D-70569 Stuttgart, Germany}
%
%
\begin{abstract}
GdN bulk is studied with the local
density approximation, on the Hartree-Fock level, and on the level
of the hybrid functional B3LYP. A local basis set formalism is used,
as implemented in the present CRYSTAL06 release. It is demonstrated
that the code is technically
capable of treating this system with its 4$f$ electrons
explicitly, i.e. out of the core.
The band structure 
at the level of the local density approximation
is in good agreement with earlier 
calculations and is found to be half-metallic. The Hartree-Fock
band structure is insulating with a large gap. Interestingly, three
solutions were found at the B3LYP level. The lowest of them is insulating
for majority spin, and the Fermi surface for minority spin
consists only of points, resulting in a very low density of states around
the Fermi level.
\end{abstract}
\submitto{\JPCM}
\maketitle 
\section{\label{Intro}Introduction}
GdN crystallizes in a rock salt structure and is by now established
as a ferromagnet with a Curie temperature of $\sim$ 69 K 
\cite{Busch1967,Li1994,Khazen2006,Granville2006}. The question
whether it is an insulator or not is still under discussion. 
In the transmittance
spectrum, a gap of 1 eV was observed \cite{Schneemeyer1987}.
From optical reflection and derivation of the plasma resonance
and from the Hall effect\cite{Wachter1980} it was argued that 
GdN was a semimetal. Thin GdN layers were found to be insulating in
resistivity measurements \cite{Xiao1996}. In X-ray photoelectron
spectroscopy (XPS) experiments, the occupied 4$f$ bands were found at a
binding energy of 7.8 eV for GdN \cite{Leuenberger2005}. 
For the related systems GdX
(X=P, As, Sb, Bi), the occupied Gd 4$f$ states were found at $\sim$ 9 eV,
and the unoccupied states at $\sim$ 5 eV in XPS and
X-ray bremsstrahlung isochromat spectroscopy (X-BIS) measurements
\cite{Yamada1996}.

The first theoretical study of GdN was an augmented plane wave calculation
using the Slater exchange potential \cite{Slater1951},
which resulted in an insulator with a very small gap \cite{Hasegawa1977}.
The $f$ electrons were treated as core states in this calculation.
A further calculation with the $f$ electrons in core \cite{Petukhov1996} 
using the 
local density approximation (LDA) gave a similar band structure,
but without gap. The spin-polarized solution was found to be metallic
for majority, and insulating for minority spin. Recently, several
calculations with explicitly treating the $f$ electrons were performed.
When the pure LDA was used, a metal was obtained 
\cite{Larson2006,Antonov2007}, 
with the occupied $f$ states at $\sim$ -4 to -3 eV 
below the Fermi level. To take
into account the strongly correlated nature of the $f$ electrons,
a Hubbard $U_f$ term was applied to these states
\cite{Ghosh2005,Larson2006,Antonov2007}. This still resulted
in a metallic solution, and only when also a $U_d$ term was applied to 
the Gd $d$ states, an insulator was obtained \cite{Larson2006}. Similarly,
in \cite{Ghosh2005}, LDA+U calculations gave a metallic ground state, and by
applying further rigid shifts to the 5$d$ and 4$f$ states, an insulating
state was found.
When
the lattice was expanded, also a transition to an
insulator was predicted, at the LDA+U level \cite{Duan2005}. A slightly
different approach to take into account the strongly correlated nature
of this system is to apply a self-interaction correction. This led again
to a half-metallic ground state for GdN, with a gap in the minority
channel \cite{Aerts2004}.

Therefore, there are many reasons to study GdN. Besides these
aforementioned reasons, from the
theory point of view, a further point is that it would
be interesting to test the performance of hybrid functionals, as a
further method to treat strongly correlated systems. It has been
shown that they give surprisingly good values for the band gaps for
such systems, where the local density approximation and other
standard functionals often fail \cite{Bredow,Iberio,Joe}. Recently, the
B3LYP hybrid functional was applied to UO$_2$, with explicitly treated
$f$ electrons, and also a good value for the gap was obtained \cite{UO2}. 
Also, plutonium oxides were found to be well described by hybrid
functionals \cite{PuO}. Cerium oxides were
studied using screened hybrid density
functionals with a local basis set\cite{Hay2006}, and with a plane wave
basis set and hybrid functionals \cite{DaSilva2007}.

The CRYSTAL code used in the present work is based on a local Gaussian
basis set. The first version released in 1988 was a pure Hartree-Fock code.
Hybrid functionals, which use an admixture of exact (Fock) exchange,
were available from the 1998 release onwards. It turned out that,
in the CRYSTAL implementation, the CPU
times for calculations with hybrid functionals are comparable to the CPU times
for calculations with standard functionals such as the LDA or gradient
corrected functionals, 
and also the memory requirements are similar in both cases.

The present CRYSTAL release \cite{Manual06} was announced of being able to
use $f$-functions as polarization functions. In this article, it will
be shown that it is also technically possible to perform
calculations on systems with $f$ electrons, with GdN as an example.
There are various reasons why GdN was chosen as a test case: firstly, the 
$f$-occupancy if $f^7$, and the half-filled $f$-shell is probably one of
the easiest examples to start with calculations on $f$-electron systems.
Also, the crystal structure is fairly simple (NaCl type).
In addition, as mentioned earlier, 
the system has already been theoretically described with other codes. 
This gives the opportunity to compare the present
calculations, where possible, with the results obtained earlier.

\section{\label{Meth} Method and calculational details}
%
%

The calculations were performed with the CRYSTAL06 code \cite{Manual06}
which uses a local basis set. GdN was considered in the face centered
cubic lattice. A ferromagnetic order was assumed, as this is
the experimentally observed state.

The local density approximation,
the Hartree-Fock approach (HF), and the B3LYP hybrid functional were
employed. The local basis sets were chosen as follows: for Gd, 
a small core pseudopotential \cite{Dolg1989} was used, which
includes scalar-relativistic effects (mass-velocity, Darwin 
and averaged spin-orbit operator \cite{WoodBoring1978}).
Together with the pseudopotential,
the corresponding basis set was
used \cite{Cao2002}, with the following modification: 
the inner $[8s7p4d4f]$ were kept as in the original basis set, and in addition
one diffuse $s$, $p$ and $d$ function with exponent 0.12 were added,
so that a $[9s8p5d4f]$ basis set was obtained. The $[3s2p1d]$ 
nitrogen basis set from reference \cite{Urea1990} was used (6-21G as in
\cite{binkley1980} and a $d$-function with exponent 0.8). The present
release of CRYSTAL06 can not compute the atomic solution of atoms
with occupied $f$-orbitals. This solution would normally be used as an initial
guess for
the subsequent calculations on the periodic system. To overcome this
problem, the atomic occupancy of Gd was 
chosen as $4s^24p^64d^{10}5s^25p^65d^86s^2$
which is obviously unusual, but sufficient to overcome the problem
of having $f$-orbitals occupied ($1s^22s^22p^63s^23p^63d^{10}$ electrons
are described
by the pseudopotential). For nitrogen, the  $1s^22s^22p^3$ state was
occupied. No convergence is achieved for the Gd atom 
which is not too surprising due to the unusual occupancy, but this
does not cause severe problems either. To converge GdN bulk, 
one convergence strategy is then e.g. to use the Anderson
mixing scheme \cite{Andersonmix}
 with 80\% mixing, and in addition to keep the spin locked
to a value of $7/2$ in the first few (e.g. 10) cycles. In some cases,
slight variations of this approach were necessary, but in the end convergence
could be achieved in all cases. The cases of LDA and HF are more
straightforward, whereas B3LYP has the additional difficulty that three
different solutions were found. To compute the corresponding potential
curves and determine the minimum, one certain solution was used as the
initial guess for the other geometries. This way, the
potential curves on the B3LYP level could be computed for all three solutions
found.

A sampling with
16 $\times$ 16 $\times$ 16 $\vec k$-points in the reciprocal lattice was used.
In the case of B3LYP, calculations without smearing and with a smearing
of 0.01 $E_h$ ($E_h$ are hartree units, 1 $E_h$=27.2114 eV)
were performed, to explore possible changes due to the
different decay properties of the density matrix in metallic
systems \cite{GoedeckerRMP}.
However, no severe differences between the calculations with and without
smearing temperature were observed.

The remaining parameters such as grids for the numerical density functional
integration were chosen as the default. The only parameter which needed
to be varied were the various thresholds for integral selection
\cite{ITOLfootnote}.

\begin{table*}
\caption{\label{Groundstateproperties}The computed equilibrium lattice
  constant and bulk modulus of GdN, at various levels of theory.}
\begin{indented}
\lineup
\item[]
\begin{tabular}{lccccc}
\br
	& $a$  & $B$   \\
	& (\AA)&   (GPa)        \\ \hline
HF & 5.10 & 167  \\
LDA & 4.91 & 174 \\
LDA (ref. \cite{Petukhov1996}), 4$f$ in core  & 4.999 & 188 \\
LDA (ref. \cite{Larson2006}), 4$f$ explicit & 4.98 & \\
B3LYP (solution with lowest energy) & 5.10 & 137 \\ 
exp. & 4.99 \cite{Klemm1956}&  192 $\pm$ 35 \cite{McWhan} \\
\br
\end{tabular}
\end{indented}
\end{table*}

\section{\label{bandstructure}Results}
\subsection{LDA calculations}

\begin{figure}
\begin{center}
\includegraphics[width=12cm,angle=270]{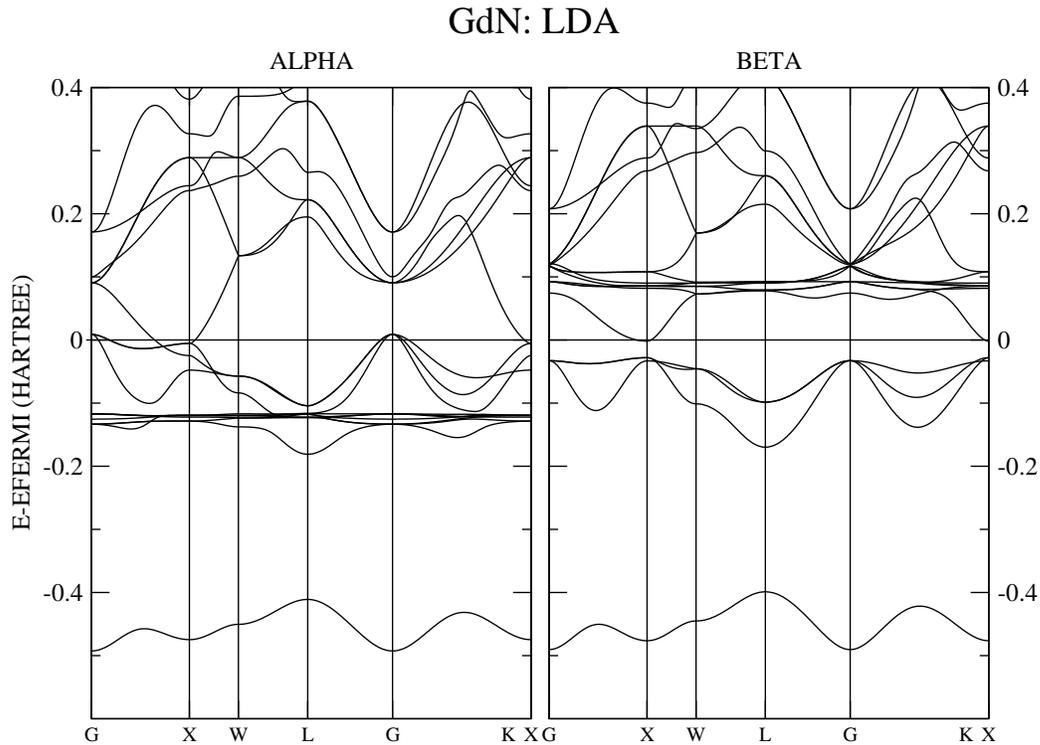}
\end{center}
\caption{\label{LDAbandstructure} LDA band structure of GdN at the computed
equilibrium lattice constant of 4.91 \AA. The Fermi energy is positioned
at 0. }
\end{figure}

The computed LDA equilibrium lattice constant of 4.91 \AA \ and
the bulk modulus of 174 GPa (table \ref{Groundstateproperties}) agree
reasonably well with the literature.
The LDA band structure is displayed in figure \ref{LDAbandstructure}.
It is half-metallic, because the bands for majority spin cross the Fermi
energy, whereas the minority bands have a gap.
The $f$-bands are positioned at $\sim$ -3.1 eV
below and $\sim$ 2.6 eV above the Fermi energy. The remaining
bands are as follows: the lowest
band displayed is N $s$, followed by three N $p$ bands which hybridize
with the Gd $d$ bands which are next higher in energy.
The band 
structure is in good agreement with the recent 
linearized muffin tin orbital approach \cite{Larson2006,Antonov2007}. 

The Mulliken
population analysis gives a charge transfer of -0.7 to the nitrogen
atoms. The Gd $f$
population is 7.1, the 5$d$ population is 1.3, and the 6$s$ and 6$p$
populations are 0.4 each. 
The spin population is 7.1 on Gd (6.9 due to the $f$ electrons
and 0.2 due to the $d$ electrons) and -0.1 on N (due to the $p$ states), which
agrees well with earlier LDA calculation where the $f$-states were
treated in core \cite{Petukhov1996}. 

\begin{figure}
\begin{center}
\includegraphics[width=12cm,angle=270]{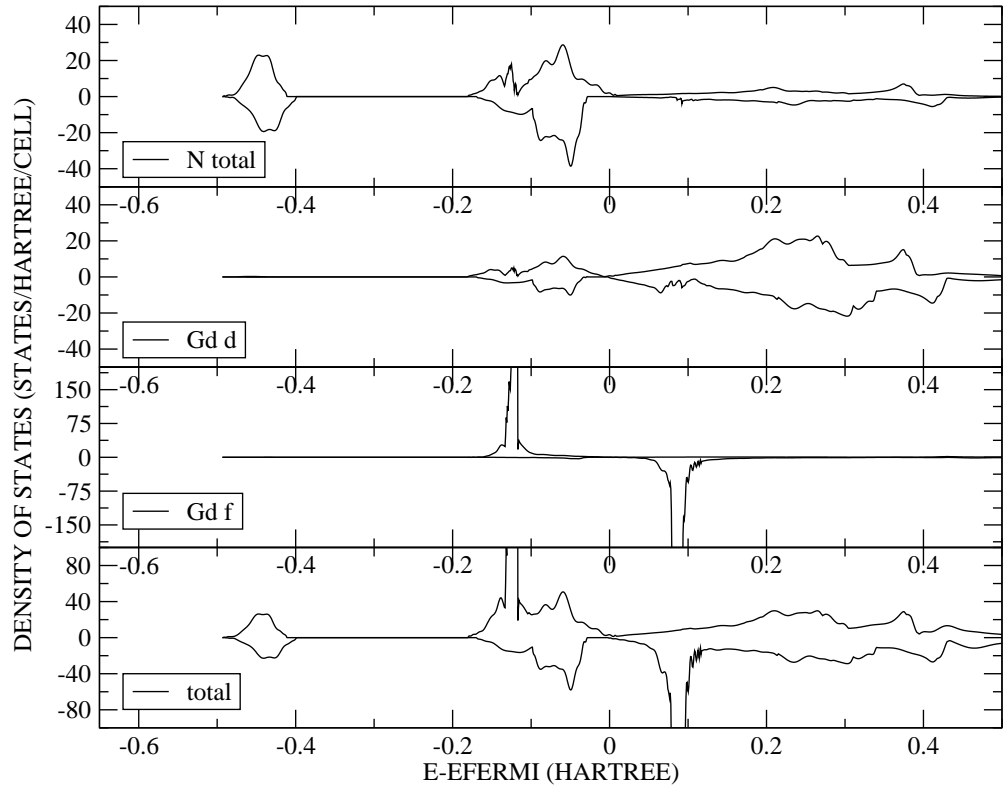}
\end{center}
\caption{\label{LDAdos} LDA density of states at the computed
equilibrium lattice constant of 4.91 \AA. Besides the total density
of states, the projected density
of states is shown for projections on N, Gd d, Gd f states.
The Fermi energy is positioned at 0. The upper part of
each panel is for majority spin, the lower part for minority spin.}
\end{figure}

The corresponding density of states is displayed
in figure \ref{LDAdos}. In the
selected energy range, the mainly occupied states are 
N ($s$ and $p$), Gd $d$ and Gd $f$. The 
density of states, projected on these states is also shown. 

%

\subsection{HF calculations}

\begin{figure}
\begin{center}
\includegraphics[width=12cm,angle=270]{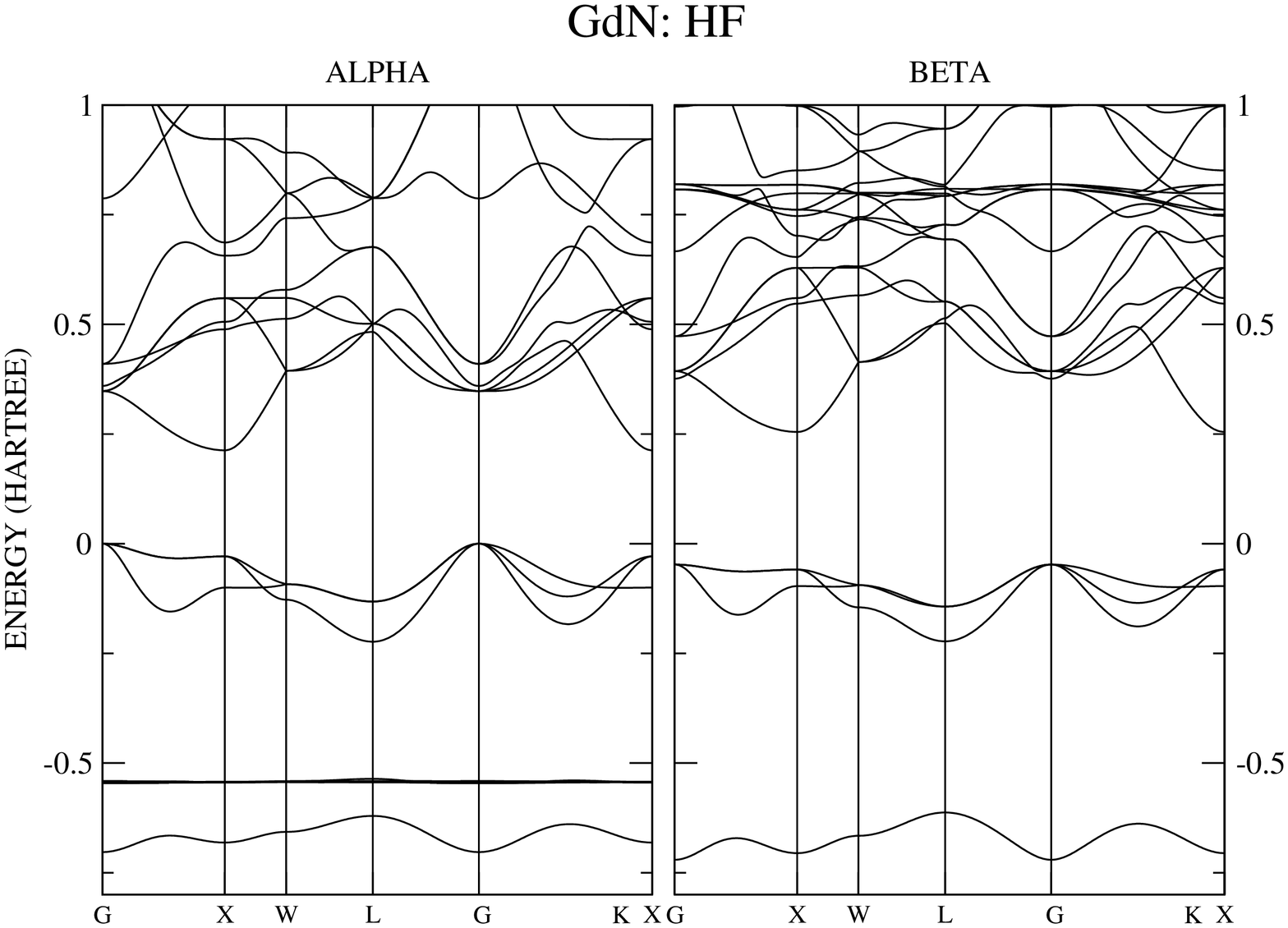}
\end{center}
\caption{\label{HFband} HF band structure of GdN at the computed
equilibrium lattice constant of 5.10 \AA. The top of the valence band
is positioned at 0.}
\end{figure}

\begin{figure}
\begin{center}
\includegraphics[width=12cm,angle=270]{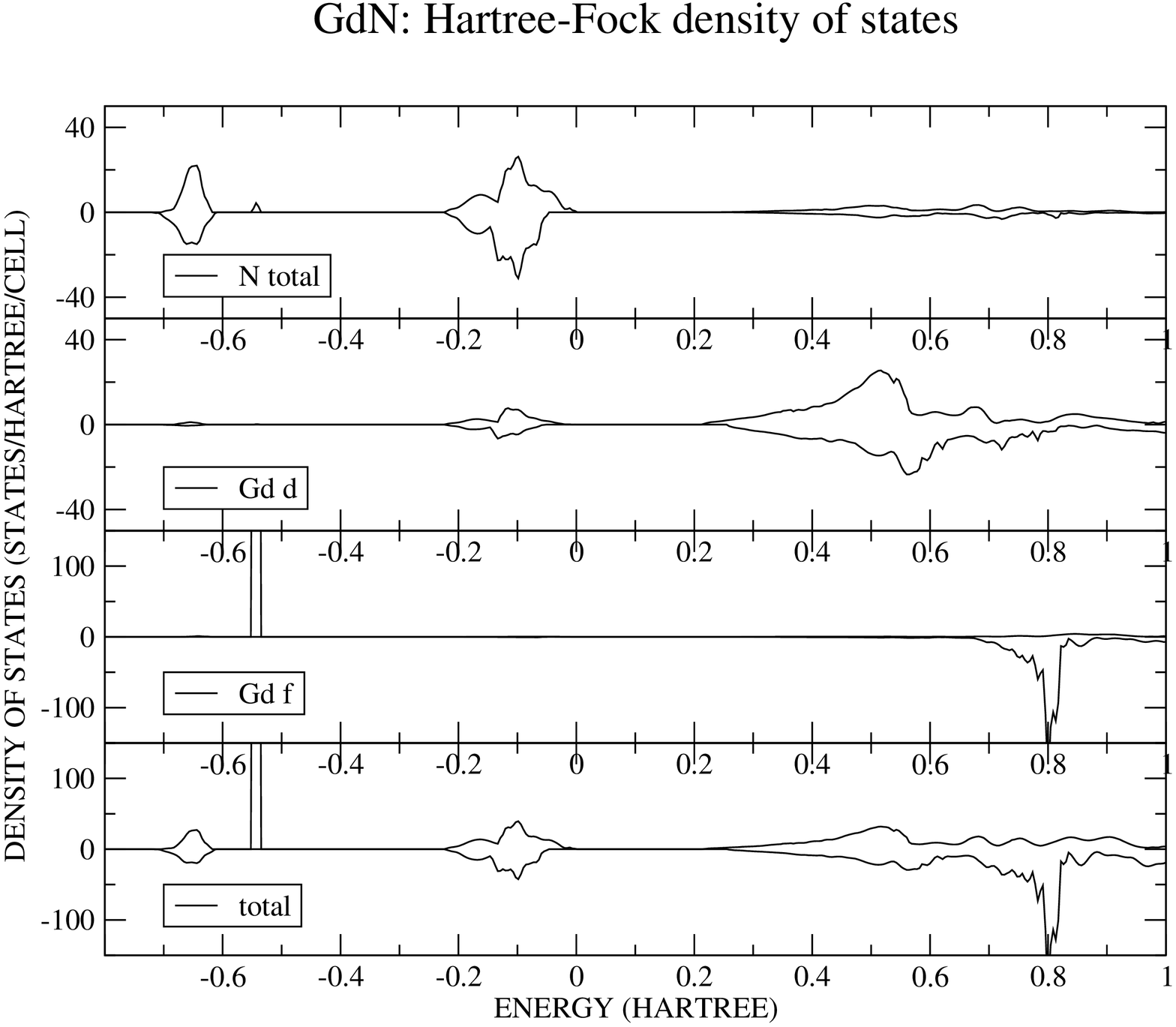}
\end{center}
\caption{\label{HFdos} HF density of states at the computed
equilibrium lattice constant of 5.10 \AA. Besides the total density
of states, the projected density
of states is shown for projections on N, Gd d, Gd f states.
The top of the valence band is positioned at 0.}
\end{figure}

The Hartree-Fock equilibrium lattice constant is 5.10 \AA \ and thus slightly
larger than the experimental value, which was also observed in an earlier
Hartree-Fock calculation where the 4$f$ electrons were treated as core
electrons\cite{Kalvoda98}. Similarly, the bulk modulus
is slightly too small due to this overestimation of the lattice constant.
The band structure at the Hartree-Fock level is displayed in figure
\ref{HFband}. GdN is insulating on the HF level, with a gap of $\sim$ 5
eV. The occupied $f$ bands are pushed downwards to $\sim -15$ eV below the
top of the valence bands, and similarly the unoccupied $f$ bands are
pushed upwards to $\sim$ 22 eV. This is to be expected, since any screening,
which is present on the LDA level, is missing at the HF level, and thus
the localized $f$ electrons feel the full, unscreened Coulomb repulsion.
Another striking feature is that the N $p$ bands and the Gd $d$ bands
are now well separated. This is also visible in the corresponding
density of states in figure \ref{HFdos}.

The total Mulliken population of N is -1.1 $|e|$. The spin population is
7.2 on Gd (7.0 due to the Gd $f$ electrons, and the remaining 0.2 mainly
due to the Gd $d$ states), and -0.2 for N (practically completely due to
the N $p$ states).

\section{B3LYP calculations}

\begin{figure}
\begin{center}
\includegraphics[width=12cm,angle=270]{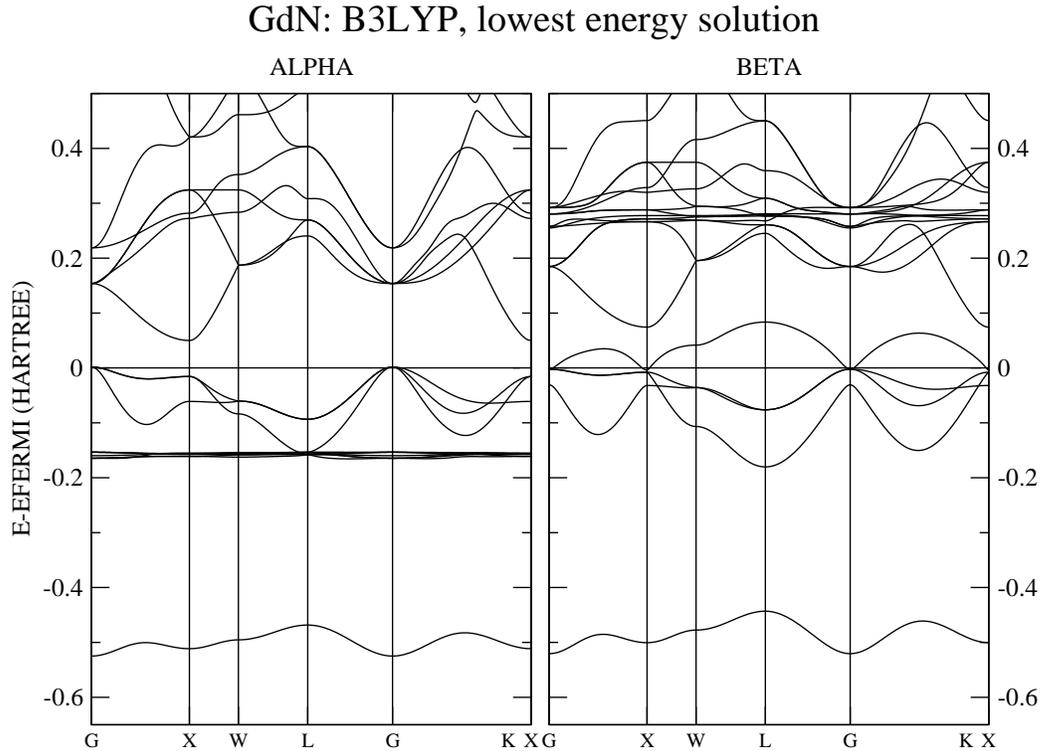}
\end{center}
\caption{\label{B3LYPbandlowest} B3LYP band structure of GdN at the computed
equilibrium lattice constant of 5.10 \AA, for the energetically most
favorable solution. The Fermi energy
is positioned at 0.}
\end{figure}

\begin{figure}
\begin{center}
\includegraphics[width=12cm,angle=270]{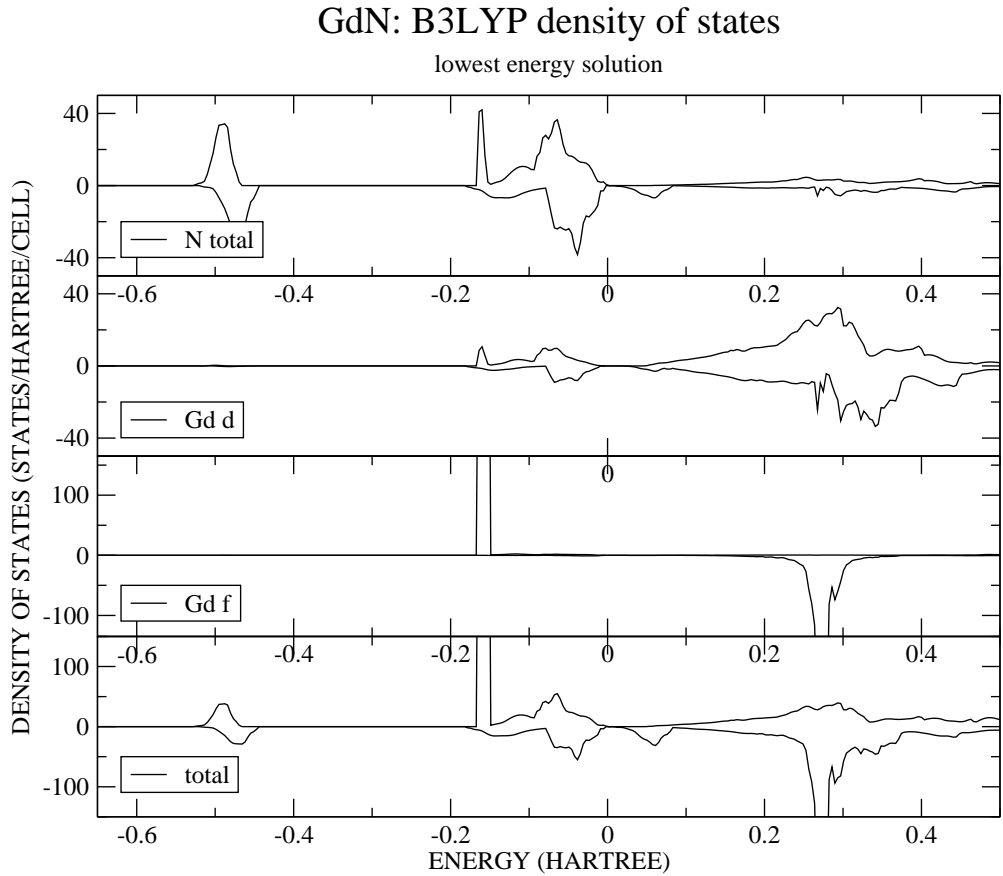}
\end{center}
\caption{\label{B3LYPdoslowest} B3LYP density of states at the computed
equilibrium lattice constant of 5.10 \AA, for the energetically most
favorable solution. Besides the total density
of states, the projected density
of states is shown for projections on N, Gd d, Gd f states.
The Fermi energy is positioned at 0.}
\end{figure}

The B3LYP calculations provided the somewhat surprising result that
two states were found, which are close in energy. The state with
the lowest energy will be discussed first. The computed lattice constant
of this state is 5.10 \AA. Its band structure is 
displayed in figure \ref{B3LYPbandlowest}, and the corresponding
density of states in figure \ref{B3LYPdoslowest}. The $f$ bands 
are positioned at $\sim$ -5 eV below and $\sim$ 7 eV above the Fermi energy,
which agrees better with the experiment, compared to LDA and HF.
In this case, the majority bands have a gap, in contrast to the LDA.
The minority valence and conduction bands only touch in various places,
so that the density of states at the top of the valence band practically
vanishes.

The N population is -0.8 $|e|$. The Gd spin population is 6.8 
(7.0 due to the $f$ electrons, 0.1 due to the $d$ electrons
and -0.3 due to the Gd $s$ basis functions),
and the N spin population is 0.2. 

\begin{figure}
\begin{center}
\includegraphics[width=12cm,angle=270]{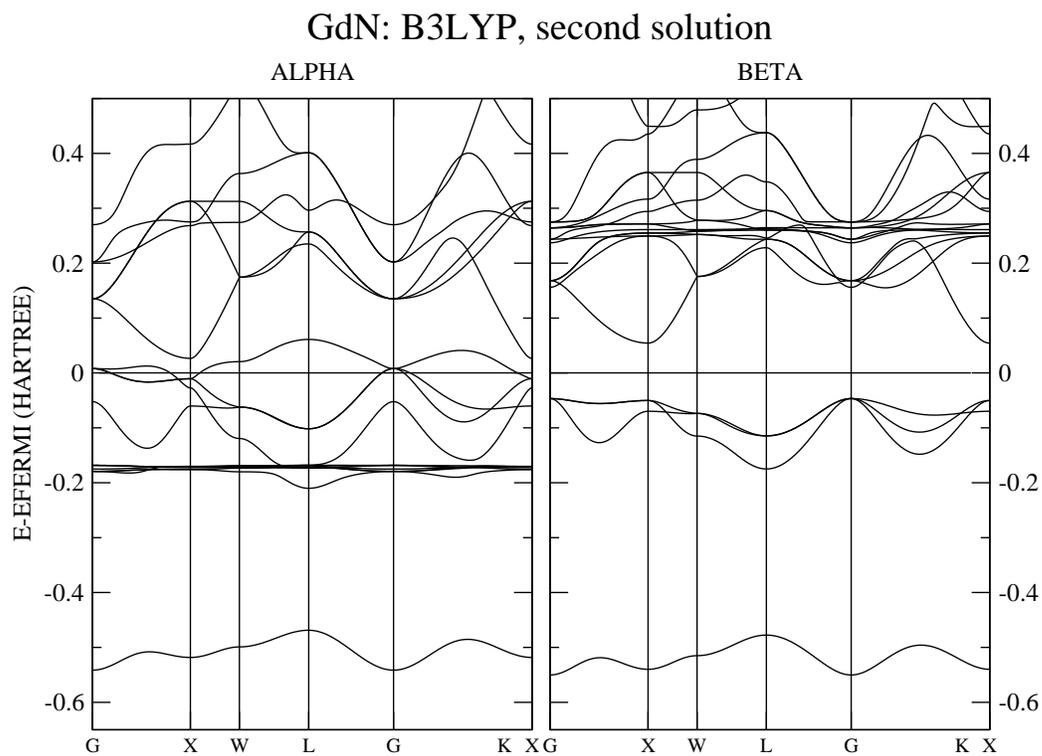}
\end{center}
\caption{\label{B3LYPband2nd} B3LYP band structure of GdN at the computed
equilibrium lattice constant of 5.10 \AA. This solution is slightly higher
(0.1 eV)
in energy than the solution in figure \ref{B3LYPbandlowest}. The Fermi energy
is positioned at 0.}
\end{figure}

\begin{figure}
\begin{center}
\includegraphics[width=12cm,angle=270]{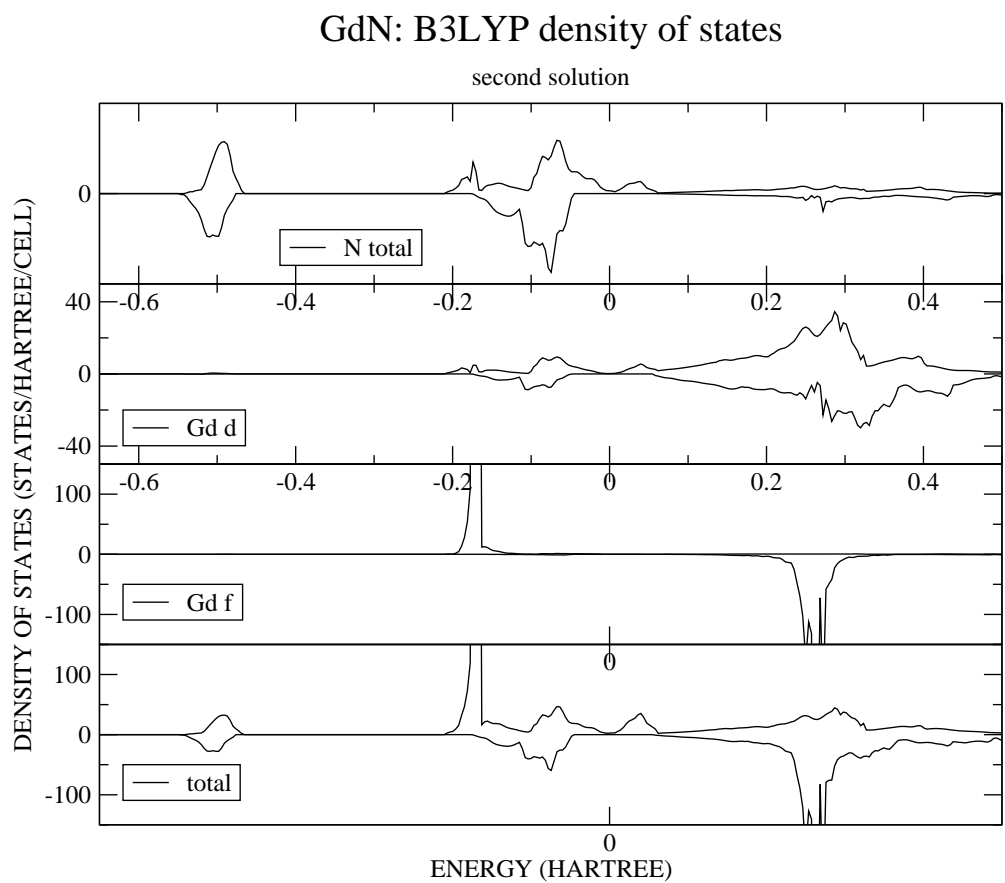}
\end{center}
\caption{\label{B3LYPdos2nd} B3LYP density of states at the computed
equilibrium lattice constant of 5.10 \AA. This solution is slightly (0.1 eV)
higher in energy than the lowest one in figure \ref{B3LYPdoslowest}. 
Besides the total density
of states, the projected density
of states is shown for projections on N, Gd d, Gd f states.
The Fermi energy is positioned at 0.}
\end{figure}

The second state is 0.004 $E_h$  (0.1 eV) higher in energy, with a
minimum at 5.11 \AA, and a bulk modulus of 133 GPa. Its band
structure is displayed in figure \ref{B3LYPband2nd}, and the
corresponding density of states in figure \ref{B3LYPdos2nd}.
The $f$ bands are similarly to those of the first state (the one
lowest in energy)
positioned at $\sim$ -5 eV below and $\sim$ 7 eV above the Fermi level. 
This state is half-metallic, in the same way as the LDA solution: 
the majority bands do not have a gap,
whereas the minority bands have a gap. 

The N population is also -0.8 $|e|$. The Gd spin population is
7.4 (7.0 due to the $f$ electrons, 0.1 due to the $d$ electron and
0.3 due to the Gd $s$ basis functions). The N spin is -0.4 and thus
antiparallel to the total Gd spin, as in the LDA. 

The fact that there are two solutions close is energy
is apparently due to the large
overlap of the $f$ bands
with the mainly nitrogen bands, especially at the L point where the nitrogen
band minimum coincides with the Gd $f$ bands. This overlap also
explains why the spin populations are a bit different from the LDA and HF
values.
Having the nitrogen spin
parallel or antiparallel does apparently not make a huge energy difference at
the B3LYP level, and these states are close in energy.

\begin{figure}
\begin{center}
\includegraphics[width=12cm,angle=270]{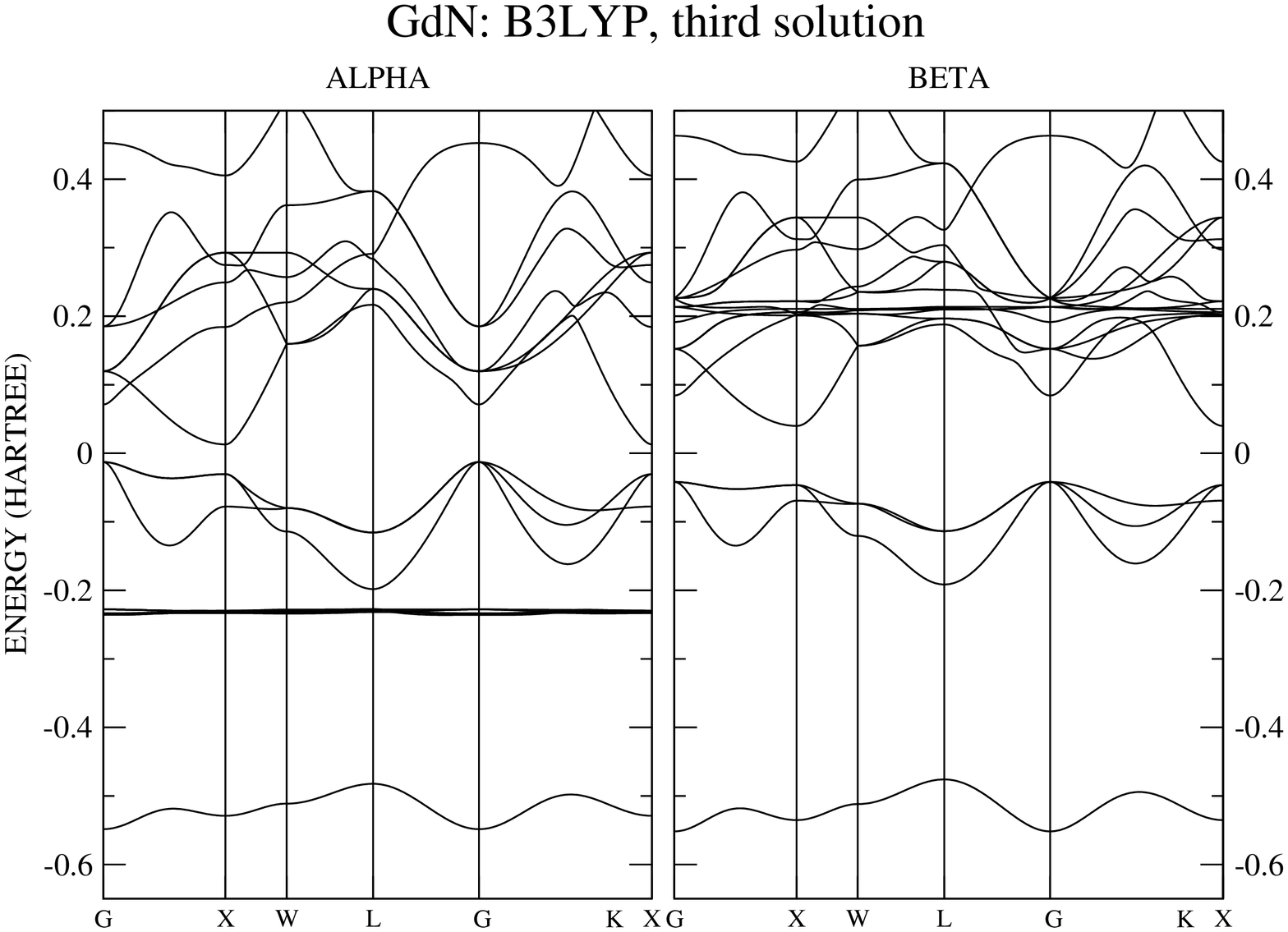}
\end{center}
\caption{\label{B3LYPbandthird} 
B3LYP band structure of GdN at the 
lattice constant of 5.10 \AA, for the insulating solution. 
This solution is the highest found at the B3LYP level, about 2.4 eV higher
than the energy of the lowest solution in figure \ref{B3LYPbandlowest}. 
The middle of the gap is positioned at 0.}
\end{figure}

Even a third solution is found, with its band structure displayed in
figure \ref{B3LYPbandthird}. 
This solution is however 0.089 $E_h$ ($\sim$ 2.4 eV)
higher in energy than the lowest solution (the equilibrium lattice constant
would be slightly shorter, 5.05 \AA, 
which does however not have a huge
impact, and the properties of this state at 5.10 \AA \ are similar). 
The computed bulk modulus of this solution is 159 GPa.
Still, it is interesting to
study this solution: the $f$ bands are now separated from the nitrogen
bands, and this solution is insulating, with a gap of $\sim$ 0.7 eV.
The charge transfer to nitrogen is $\sim$ -0.9, and the spin is
now more similar to the LDA and HF solutions: 7.1 for Gd in total
(7.0 due to the $f$ electrons, 0.1 due to the $d$ electrons), and
-0.1 for N due to the $p$ electrons. This confirms that the somewhat
unusual population for the two lowest states on the B3LYP level are
due to the hybridization at the L point. This third solution has also
the property of "interpolating" between LDA and HF (lattice constant,
shape of the band structure), what is often observed
for
the B3LYP hybrid functional: due to the admixture of Fock exchange, the
results for B3LYP calculations are usually expected to be in the range
in between LDA and HF.

%

\newpage
\section{Conclusion}
It was demonstrated that Gaussian type orbitals are technically capable
for performing calculations
on GdN bulk with $f$ electrons explicitly treated. The LDA results are
in very good agreement with previous calculations and give a half-metallic
solution, with the majority bands being conducting and the minority
bands insulating. The Hartree-Fock solution is insulating with
a large gap of $\sim$ 5 eV, which is a typical overestimation of the gap
at the Hartree-Fock level due to the lack of screening. On the B3LYP
level, three solutions were found. The lowest one has a gap for
majority spin; and for minority spin, 
the valence and conduction bands only touch at
certain points of the Brillouin zone. The corresponding density
of states is thus very small around the Fermi energy. The second solution
is very close in energy ($\sim$ 0.1 eV higher), and the 
majority bands cross the Fermi energy. A third solution was found to
be insulating, but this third solution is about 2.4 eV higher in energy
than the lowest solution. 
The fact that there are two
nearly degenerate solutions makes the comparison with the experiment
very difficult, and it is difficult to judge the performance of
the hybrid functional B3LYP in this case. 
It is however also very interesting that two
different solutions with very different Fermi surfaces are obtained
in the calculations.
As the experimental situation is not fully clear,
experiments such as photoemission on this system might be very interesting
to further elucidate the electronic structure of GdN
(indeed, spin and angle resolved inverse photoemission spectroscopy was also
suggested in another recent theoretical work \cite{Sharma2006}).

\section{Acknowledgment}
This work was supported by the MMM-initiative of the Max-Planck society.

\clearpage

\bibliographystyle{unsrt}
\bibliography{gdn_resub.bib}

\begin{thebibliography}{10}

\bibitem{Busch1967}
G.~Busch.
\newblock {\em J. Appl.\ Phys.}, 38:1386, 1967.

\bibitem{Li1994}
D.~X. Li, Y.~Haga, H.~Shida, and T.~Suzuki.
\newblock {\em Physica B}, 199-200:631, 1994.

\bibitem{Khazen2006}
K.~Khazen, H.~J. von Bardeleben, J.~L. Cantin, A.~Bittar, S.~Granville, H.~J.
  Trodahl, and B.~J. Ruck.
\newblock {\em Phys.\ Rev.\ B}, 74:245330, 2006.

\bibitem{Granville2006}
S.~Granville, B.~J. Ruck, F.~Budde, A.~Koo, D.~J. Pringle, F.~Kuchler, A.~R.~H.
  Preston, D.~H. Housden, N.~Lund, A.~Bittar, G.~V.~M. Williams, and H.~J.
  Trodahl.
\newblock {\em Phys.\ Rev.\ B}, 73:235335, 2006.

\bibitem{Schneemeyer1987}
L.~F. Schneemeyer, R.~B. van Dover, and E.~M. Gyorgy.
\newblock {\em J. Appl.\ Phys.}, 61:3543, 1987.

\bibitem{Wachter1980}
P.~Wachter and E.~Kaldis.
\newblock {\em Solid State Commun.}, 34:241, 1980.

\bibitem{Xiao1996}
J.~Q. Xiao and C.~L. Chien.
\newblock {\em Phys.\ Rev.\ Lett.}, 76:1727, 1996.

\bibitem{Leuenberger2005}
F.~Leuenberger, A.~Parge, W.~Felsch, K.~Fauth, and M.~Hessler.
\newblock {\em Phys.\ Rev.\ B}, 72:014427, 2005.

\bibitem{Yamada1996}
H.~Yamada, T.~Fukawa, T.~Muro, Y.~Tanaka, S.~Imada, S.~Suga, D.-X. Li, and
  T.~Suzuki.
\newblock {\em J. Phys.\ Soc.\ Japan}, 65:1000, 1996.

\bibitem{Slater1951}
J.~C. Slater.
\newblock {\em Phys.\ Rev.}, 81:385, 1951.

\bibitem{Hasegawa1977}
A.~Hasegawa and A.~Yanase.
\newblock {\em J. Phys.\ Soc.\ Japan}, 42:492, 1977.

\bibitem{Petukhov1996}
A.~G. Petukhov, W.~R.~L. Lambrecht, and B.~Segall.
\newblock {\em Phys.\ Rev.\ B}, 53:4324, 1996.

\bibitem{Larson2006}
P.~Larson and W.~R.~L. Lambrecht.
\newblock {\em Phys.\ Rev.\ B}, 74:085108, 2006.

\bibitem{Antonov2007}
V.~N. Antonov, B.~N. Harmon, A.~N. Yaresko, and A.~P. Shpak.
\newblock {\em Phys.\ Rev.\ B}, 75:184422, 2007.

\bibitem{Ghosh2005}
D.~B. Ghosh, M.~Me, and S.~K. De.
\newblock {\em Phys.\ Rev.\ B}, 72:045140, 2005.

\bibitem{Duan2005}
C.~g.~Duan, R.~F. Sabiryanov, J.~Liu, W.~N. Mei, P.~A. Dowben, and J.~R. Hardy.
\newblock {\em Phys.\ Rev.\ Lett.}, 94:237201, 2005.

\bibitem{Aerts2004}
C.~M. Aerts, P.~Strange, M.~Horne, W.~Temmerman, Z.~Szotek, and A.~Svane.
\newblock {\em Phys.\ Rev.\ B}, 69:045115, 2004.

\bibitem{Bredow}
T.~Bredow and A.~R. Gerson.
\newblock {\em Phys.\ Rev.\ B}, 61:5194, 2000.

\bibitem{Iberio}
I.~de~P.~R.~Moreira, F.~Illas, and R.~L. Martin.
\newblock {\em Phys.\ Rev.\ B}, 65:155102, 2002.

\bibitem{Joe}
J.~Muscat, A.~Wander, and N.~M. Harrison.
\newblock {\em Chem.\ Phys.\ Lett.}, 342:397, 2001.

\bibitem{UO2}
K.~N. Kudin, G.~E. Scuseria, and R.~L. Martin.
\newblock {\em Phys.\ Rev.\ Lett.}, 89:266402, 2002.

\bibitem{PuO}
I.~D. Prodan, G.~E. Scuseria, J.~A. Sorda, K.~N. Kudin, and R.~L. Martin.
\newblock {\em The Journal of Chemical Physics}, 123:014703, 2005.

\bibitem{Hay2006}
P.~J. Hay, R.~L. Martin, J.~Uddin, and G.~E. Scuseria.
\newblock {\em The Journal of Chemical Physics}, 125:034712, 2006.

\bibitem{DaSilva2007}
J.~L. F.~Da Silva, M.~V. Ganduglia-Pirovano, J.~Sauer, V.~Bayer, and G.~Kresse.
\newblock {\em Phys.\ Rev.\ B}, 75:045121, 2007.

\bibitem{Manual06}
{R. Dovesi, V. R. Saunders, C. Roetti, R. Orlando}, C.~M. Zicovich-Wilson,
  F.~Pascale, B.~Civalleri, K.~Doll, N.~M. Harrison, I.~J. Bush, Ph. D'Arco,
  and M.~Llunell.
\newblock {\em {\sc crystal 2006} User's Manual}.
\newblock University of Torino, Torino, 2006.

\bibitem{Dolg1989}
M.~Dolg, H.~Stoll, and H.~Preuss.
\newblock {\em The Journal of Chemical Physics}, 90:1730, 1989.

\bibitem{WoodBoring1978}
J.~H. Wood and A.~M. Boring.
\newblock {\em Phys.\ Rev.\ B}, 18:2701, 1978.

\bibitem{Cao2002}
X.~Cao and M.~Dolg.
\newblock {\em J. Mol. Struct. (Theochem)}, 581:139, 2002.

\bibitem{Urea1990}
R.~Dovesi, M.~Caus\`a, R.~Orlando, C.~Roetti, and V.~R. Saunders.
\newblock {\em The Journal of Chemical Physics}, 92:7402, 1990.

\bibitem{binkley1980}
J.~S. Binkley, J.~A. Pople, and W.~J. Hehre.
\newblock {\em J. Am. Chem. Soc.}, 102:939, 1980.

\bibitem{Andersonmix}
D.~G. Anderson.
\newblock {\em J. Assoc. Comput. Mach.}, 12:547, 1965.

\bibitem{GoedeckerRMP}
S.~Goedecker.
\newblock {\em Rev.\ Mod.\ Phys.}, 71:1085, 1999.

\bibitem{ITOLfootnote}
In the HF case, the values for the selection of the Coulomb integrals
  (parameter one and two) and exchange integrals (parameter three, four and
  five, as defined in the manual \cite{Manual06}) were chosen as
  $10^{-9},10^{-9},10^{-9},10^{-9},10^{-18}$, in order to avoid linear
  dependence problems. In the case of LDA and B3LYP, the default values
  ($10^{-6},10^{-6},10^{-6},10^{-6},10^{-12}$) led to numerically stable
  solutions.

\bibitem{Klemm1956}
W.~Klemm and G.~Winkelmann.
\newblock {\em Z. anorg. allg. Chem.}, 288:87, 1956.

\bibitem{McWhan}
D.~B. McWhan.
\newblock {\em The Journal of Chemical Physics}, 3528:44, 1966.

\bibitem{Kalvoda98}
S.~Kalvoda, M.~Dolg, H.-J. Flad, P.~Fulde, and H.~Stoll.
\newblock {\em Phys.\ Rev.\ B}, 57:2127, 1998.

\bibitem{Sharma2006}
A.~Sharma and W.~Nolting.
\newblock {\em J. Phys.: Condens.\ Matter}, 18:7337, 2006.

\end{thebibliography}
\end{document}